\documentclass[conference]{IEEEtran}
\IEEEoverridecommandlockouts

\usepackage{cite}
\usepackage{amsmath,amssymb,amsfonts}
\usepackage{algorithmic}
\usepackage{graphicx}
\usepackage{textcomp}
\usepackage{xcolor}

\usepackage{bm}
\usepackage{multirow}
\usepackage{tabularx}

\usepackage{url}
\usepackage{booktabs}       
\usepackage{nicefrac}       
\usepackage{microtype}      
\usepackage{caption}
\usepackage{subfigure}

\usepackage{listings}

\def\BibTeX{{\rm B\kern-.05em{\sc i\kern-.025em b}\kern-.08em
    T\kern-.1667em\lower.7ex\hbox{E}\kern-.125emX}}

\begin{document}

\title{SA-OOSC: A Multimodal LLM-Distilled Semantic Communication Framework for Enhanced Coding Efficiency with Scenario Understanding}

\author{
Feifan~Zhang$^*$, Yuyang~Du$^*$, Yifan~Xiang, Xiaoyan~Liew, Soung~Chang~Liew
\thanks{$^*$Authors contribute equally. S. C. Liew is the corresponding author.}
\thanks{F. Zhang, Y. Du, Y. Xiang, X. Liew and S. C. Liew are with the Department of Information Engineering, The Chinese University of Hong Kong, HKSAR, China (e-mail: \{zf024, yuydu, soung\}@ie.cuhk.edu.hk, 
\{xyf20040227, liuxy185\}@link.cuhk.edu.hk).}
\vspace{-2.5em}
}
\maketitle

\begin{abstract}
This paper introduces SA-OOSC, a multimodal large language models (MLLM)-distilled semantic communication framework that achieves efficient semantic coding with scenario-aware importance allocations. This approach addresses a critical limitation of existing object-oriented semantic communication (OOSC) systems -- assigning static importance values to specific classes of objects regardless of their contextual relevance. Our framework utilizes MLLMs to identify the scenario-augmented (SA) semantic importance for objects within the image. Through knowledge distillation with the MLLM-annotated data, our vectorization/de-vectorization networks and JSCC encoder/decoder learn to dynamically allocate coding resources based on contextual significance, i.e., distinguishing between high-importance objects and low-importance according to the SA scenario information of the task. The framework features three core innovations: a MLLM-guided knowledge distillation pipeline, an importance-weighted variable-length JSCC framework, and novel loss function designs that facilitate the knowledge distillation within the JSCC framework. Experimental validation demonstrates our framework’s superior coding efficiency over conventional semantic communication systems, with open-sourced MLLM-annotated and human-verified datasets established as new benchmarks for future research in semantic communications.
\end{abstract}


\section{Introduction} \label{sec1} 
With the rapid development of visual applications such as augmented reality (AR) and virtual reality (VR), as well as the growing connectivity demands of the Internet of Things (IoT), the amount of data that needs to be transmitted within a network has increased dramatically \cite{Ref000,Ref001,Ref002,llmind}. Although the evolution of wireless communication technologies from 1G to 5G has brought significant improvements in packet transmission rates, bit-based communication technologies are approaching the Shannon capacity at the physical layer \cite{arikan2009channel}. 

Semantic communication has emerged as a highly promising paradigm for next-generation wireless communication to overcome this backdrop \cite{semreview1,semreview2}. Unlike traditional communication, semantic communication focuses on transmitting the meaning of information rather than its precise representation, thus offering the potential to go beyond the capacity limitations in conventional systems \cite{semantic1,semantic2}.

Although semantic communication systems can significantly reduce bandwidth usage, their coding efficiency and efficacy can potentially be further improved. The authors in \cite{Deepjscc} proposed a pioneering joint source-channel coding (JSCC) framework that used deep neural networks (DNN) to map an image’s pixel representation into a complex-valued vector to be transmitted over the noisy channel. This method relied on a fixed rate encoding method. A subsequent advancement, referred to as NTSCC \cite{NTSCC}, introduced a variable-length JSCC scheme that divided an image into patches and estimated the information entropy of each patch with an entropy model. The entropy information was utilized to guide the variable-length JSCC process -- allocating more coding resources to regions of larger entropy value with higher texture complexity \cite{NTC2018}. While this approach marked a significant step forward, it primarily focused on the texture and structural complexity of image regions, ignoring task-relevant semantic importance that is critical for many real-world applications. To address this limitation, task-oriented semantic communication methods were developed \cite{OOSC,OOSC1,OOSC2,OOSC3}. In \cite{OOSC}, the authors proposed a reinforcement learning-based adaptive semantic coding framework that extracted semantic importance using semantic segmentation algorithms and allocated coding bits according to semantic importance -- a traffic light, for example, is considered important and is given more coding bits for autonomous driving tasks. However, these methods rely only on object segmentation to partition objects and assign predefined semantic importance values that are deterministic and independent of the specific context. Such static approaches fail to account for the dynamic nature of real-world scenarios, where the importance of the same class of object can vary depending on the practical context, leading to inefficiencies in coding resource allocation.

\begin{figure}[htbp]
\centering
\includegraphics[width=0.91\linewidth]{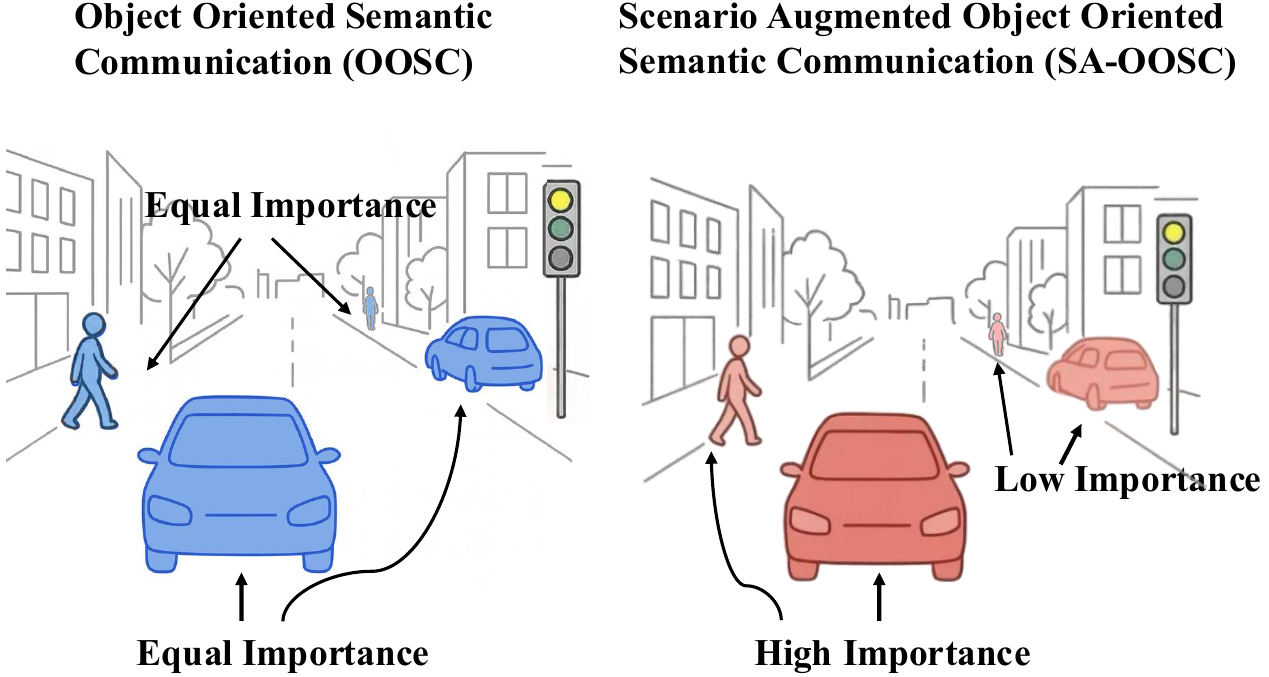} 
\captionsetup{font={small}}
\caption{Illustration of semantic importance identification in OOSC and SA-OOSC under the autonomous driving scenario. In OOSC, task-relevant objects, such as vehicles and pedestrians, are uniformly assigned equal coding importance, regardless of their contextual relevance. In SA-OOSC, an object's semantic importance is decided by the scenario understanding, prioritizing objects such as a nearby vehicle or an imminent pedestrian based on their contextual significance, to achieve more effective and adaptive semantic communication.}
    \label{fig:sa-allocation}
\end{figure}

Let us continue with the autonomous driving setting. In a vehicular network, multiple vehicles share real-time road images to build a comprehensive environmental model that supports the vehicle's decision-making \cite{V2V,V2V2,V2V3,V2V4}. As illustrated in Fig. \ref{fig:sa-allocation}, traditional task-oriented semantic communication methods -- referred as Object-Oriented Semantic Communication (OOSC) in this paper -- typically treat all objects of certain categories, such as vehicles and pedestrians, as equally important, allocating uniform coding resources to each without considering their specific locations or actual importance within the scene. This approach does not align with the cognitive and decision-making patterns of human drivers. In practice, a human driver dynamically adjusts attention based on the specific context of the scene. For example, a driver may pay greater attention to a nearby vehicle in the same driving lane than a parked vehicle at a distance. Similarly, the importance of a pedestrian crossing the road should be higher than the one walking along the road.

To overcome this limitation, we propose the Scenario Augmented OOSC (SA-OOSC) approach as depicted in Fig. \ref{fig:sa-allocation}. This method incorporates scenario understanding to analyze the semantic importance of each object according to its contextual relevance to the ongoing task. To imitate a human understanding of the task scenario, we leverage the image analysis capabilities of multimodal large language models (MLLMs) to automatically label the contextual significance of objects within the scene. In the driving scenario within Fig. \ref{fig:sa-allocation}, for example, the vehicle approaching head-on near the road centerline at close, or the nearby pedestrian crossing the road, should be treated with more attention; while the vehicle parked far away and the pedestrian walking along the sidewalk are less significant in terms of the scenario understanding.

With the massive amount of MLLM-labeled data with scenario understanding, we train the vectorization/inverse vectorization network and JSCC encoder/decoder within our semantic communication framework for knowledge distillation purposes. Our framework, referred to as the SA-OOSC framework, learns from the MLLM-labeled data in the distillation process and gains the ability to allocate more coding resources to high-importance areas (e.g., the nearby pedestrian crossing the road) and assign fewer resources to low-importance areas (e.g., the pedestrian walking along the sidewalk). As we will later show through experiments, SA-OOSC achieves more efficient and effective resource utilization than previous semantic communication systems owing to this scenario-aware importance allocation mechanism.

The rest of this paper uses autonomous driving as the validation scenario. Our major contributions are as follows: First, this paper highlights an important problem overlooked in current semantic communication systems -- dynamically determining semantic importance of different objects according to the communication scenario. To address this issue, we put forth SA-OOSC, a framework that leverages MLLM to simulate a human’s decision-making process in allocating coding resources according to an object’s semantic importance under a specific context.

Second, to leverage the ability of the MLLM, we propose a comprehensive image processing pipeline with variable-length JSCC in the implementation of SA-OOSC. This framework includes three major technical innovations: 1) knowledge distillation with MLLM-annotated data, which equips the framework with scenario understanding capabilities so that it can automatically generate the scenario-augmented semantic importance labels for objects therein; 2) assigning the corresponding importance weights to each image patch and generating their latent representations based on the labeling results; and 3) designing a novel scenario-augmented semantic loss function to optimize the variable-length JSCC scheme, thereby enabling more efficient coding resource allocation than previous methods.

Finally, as the first study to utilize MLLMs for scenario-augmented JSCC, we have open-sourced the scenario-driven semantic communication dataset annotated by GPT-4V. This dataset is positioned as a vital benchmark for training/testing future scenario-augmented semantic communication systems. Further, to validate the accuracy of GPT-4V in assigning semantic importance according to the driving scenario, we invited three experienced drivers to manually annotate a subset of the original dataset as a cross check. The valuable human-annotated data is also made publicly available and can be used to evaluate the alignment of the proposed semantic importance assignment method with human preferences.\footnote{\textcolor{blue}{\url{https://github.com/xyfyyds/Semantic-Communication-Cityscapes}}}

\section{MLLM for Knowledge Distillation} \label{sec2} 
We consider an image $S$ to be transmitted over a noisy channel, where $S \in \mathbb{R}^{h \times w \times 3}$ represents an RGB image with height $h$, width $w$, and three color channels. In our system, the MLLM-based identification module is designed to label the scenario-augmented semantic importance (henceforth referred to as the SA semantic importance) of different regions within this image.

\begin{figure}[htbp]
    \centerline{\includegraphics[width=0.7\linewidth]{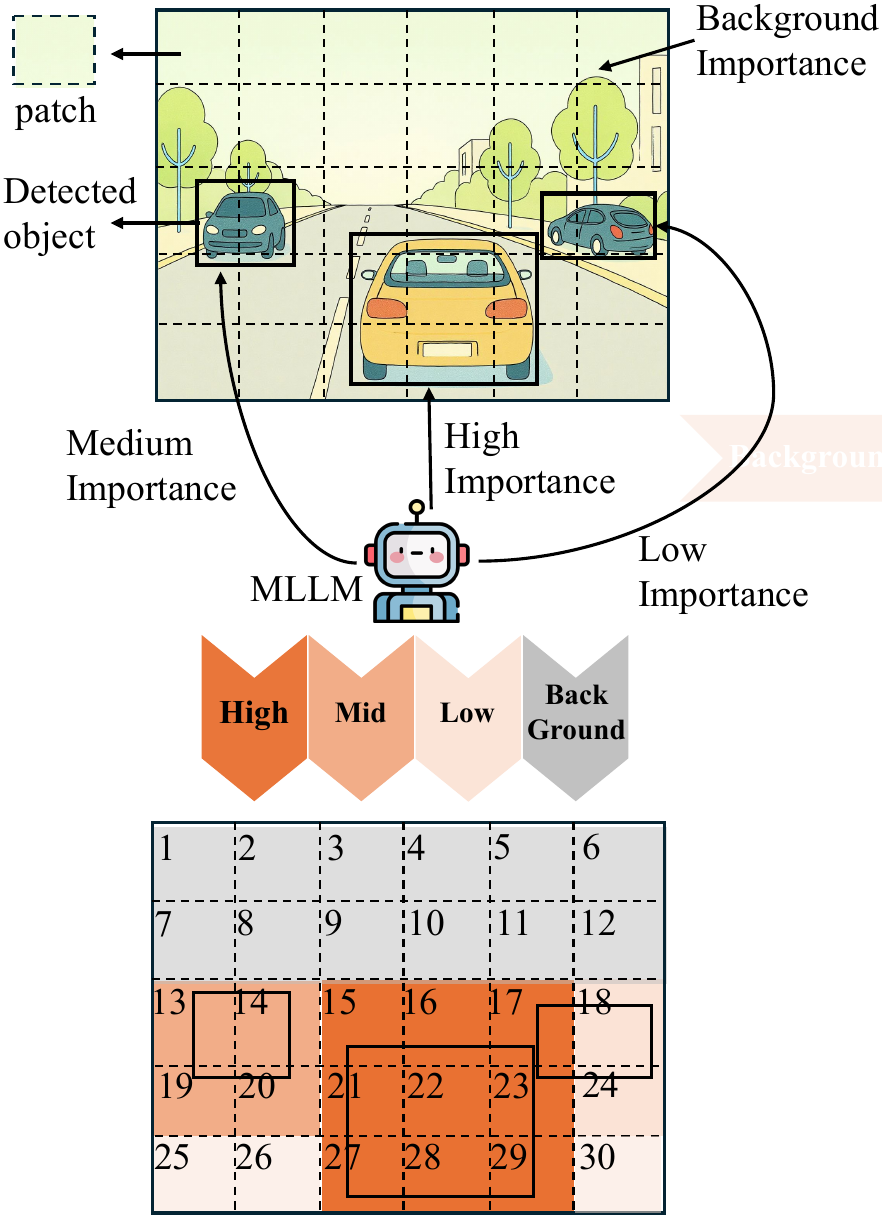}}
    \captionsetup{font={small}}
    \caption{Illustrations of the object-based (top) and the patch-based (bottom) SA semantic importance.}
    \label{fig:object_patch_assign}
\end{figure}

This module starts with an object detection process to identify and localize all objects within the input image, with identification results annotated with bounding boxes (see Fig. \ref{fig:object_patch_assign} for illustrations). In the subsequent step, the annotated image, as well as a detailed textual description of the downstream task, are given to the MLLM integrated in this module. The duty of the MLLM is to assign different levels of semantic importance to each object detected. In the autonomous driving scenario depicted in Fig. \ref{fig:object_patch_assign}, for example, the vehicle in the same lane ahead of the ego-vehicle is assigned ``high importance", the oncoming vehicle in the opposite lane at a distance is assigned ``medium importance'', while the vehicle parked at the roadside is assigned ``low importance''. Without loss of generality, we assign the importance labels of the above three types of objects (i.e., high, medium, and low importance) as $3$, $2$, and $1$, respectively.

The above assignments of SA semantic importance are object-based. Yet, image processing frameworks in semantic communication systems, including our scheme to be presented in section \ref{sec3}, are patch-based. As a typical pre-processing step before vectorization and encoding, the original image $S$ will be tokenized into $L$ non-overlapping patches, indexed in a raster-scan order (left-to-right, top-to-bottom, see patch indexes in Fig. \ref{fig:object_patch_assign} for example) to form the patch sequence $\mathbf{P} = [P_1, \ldots, P_L]$. To align with later treatments in our image semantic communication framework (see discussion in section \ref{sec3}), we must assign each image patch with an associated importance label, forming the importance sequence $\mathbf{I} = [I_1, \ldots, I_L]$ that one-on-one maps to the patch sequence $\mathbf{P}$. Therefore, we need a transformation between object-based importance (i.e., what we have obtained from the object-based importance identification, see the image on the top of Fig. \ref{fig:object_patch_assign}) and patch-based importance (i.e., what we need in later image processing, see the image at the bottom of Fig. \ref{fig:object_patch_assign}). We now explain how we assign the importance label of an image patch according to the object(s) therein:

\textbf{Single-object patch:} For an image patch containing one object only, we assign the object’s importance label to the patch. Patch 13 is an example -- the patch contains part of the medium-importance vehicle detected and is therefore identified as a medium-importance patch.

\textbf{Multiple-object patch:} For an image patch containing multiple objects, we let the patch’s importance label be the importance label of the most important object therein. One of the most typical examples is Patch 17, which contains both the low-importance vehicle parked at the roadside and the vehicle driving ahead of the ego-vehicle. The importance of Patch 17 is labeled according to the high-importance driving vehicle.

\textbf{Background patch:} For an image patch that does not contain any task-relevant object detected, such as Patch 1 to Patch 12 with far-away trees or the sky only, we assign the lowest importance to that patch. Such patches typically play non-significant roles in the downstream task. We treat them as the image’s ``background'' and assign zero, the lowest semantic importance level, to these patches.

For technical details about the MLLM-empowered semantic importance assignment process, including the deployment of the MLLM based on the GPT-4V model and the prompt engineering we have conducted, we refer readers to Appendix \ref{appendixA} for details. Here in Section \ref{sec2}, we only present the object-based importance and the patch-based importance in Fig. \ref{fig:object_patch_assign} for illustration purposes. It is worth noting that, in practical implementations, the number of patches $L$ is typically much larger than $30$ (often in the hundreds or even thousands) to facilitate more efficient encoding. Here, we use $30$ patches solely to convey the main concept.

In the following, we use the well-known Cityscapes dataset \cite{Cityscapes} as a proof-of-concept for the scenario-augmented semantic communication system. Applying the MLLM-based SA importance labeling scheme to the dataset, we obtain high-quality distillable data that incorporates the MLLM’s scenario understanding knowledge. We can use this data to train the entire networks within our semantic communication framework to equip the framework with the ability to analyze the SA semantic importance within an image. This process is known as \emph{knowledge distillation}, wherein the scenario understanding ability of a large-scale model can be transformed into smaller-scale ones with a relatively low cost \cite{KD1,KD2,wang2025rephrase,KD3}.

\section{Image Transmission within the SA-OOSC Framework}\label{sec3} 

\begin{figure*}[htbp]
    \centering
    \captionsetup{font={small}}
    \includegraphics[width=0.975\linewidth]{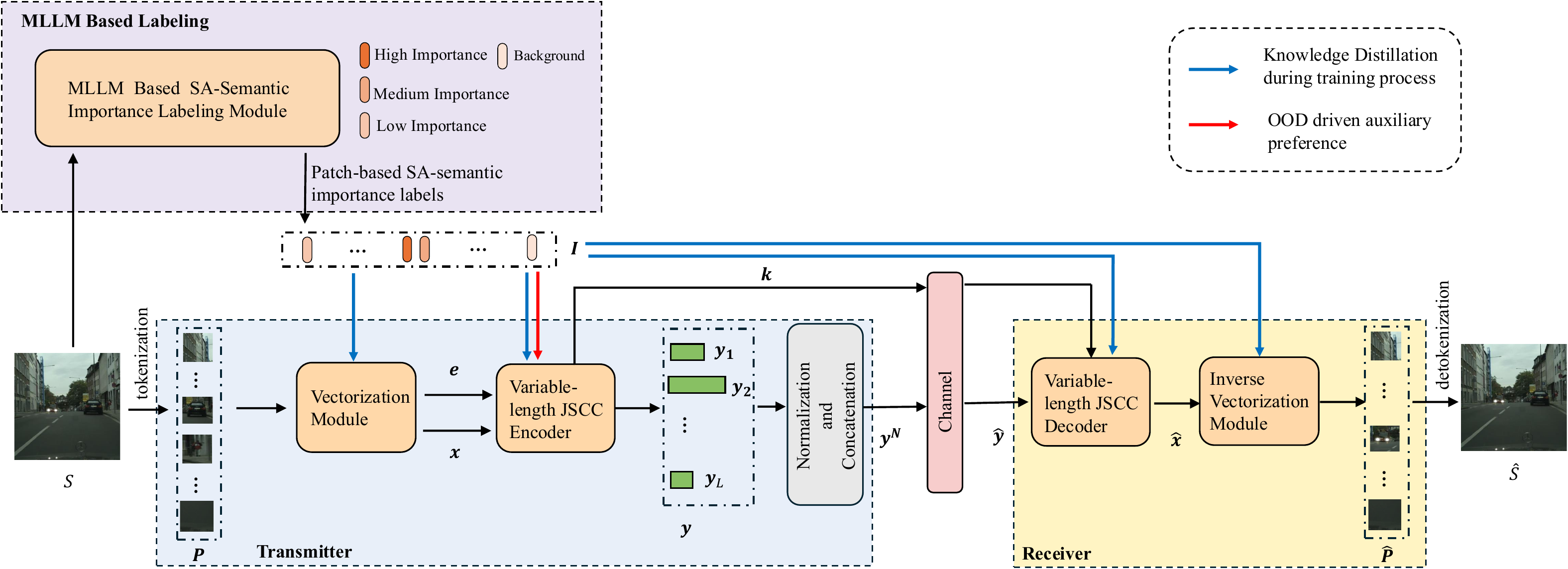} 
    \caption{The block diagram of the SA-OOSC system. The SA importance labeling module is involved in both the transmission of a tested image (red arrow) and the training of vectorization/inverse vectorization/JSCC encoding (blue arrows). Section \ref{sec3} only focuses on the signal processing pipeline for image transmissions. Model training details (i.e., related to those blue arrows) are available in Appendixes \ref{appendixB} to \ref{appendixH}.}
    \label{fig:sa-oosc-block}
\end{figure*}

This section provides a comprehensive overview of the proposed system, including a high-level description of the framework and an outline of the interaction between key modules. We present the SA-OOSC framework in an order corresponding to how an image is transmitted and recovered: first the transmitter design, then the wireless channel model, and finally the receiver design. This section focuses on the image transmission process within SA-OOSC. For model training details, such as the training procedures of each module and the loss function design therein, we refer readers to Appendixes \ref{appendixB} to \ref{appendixH} for details.

\textbf{Transmitter Design:} With reference to Fig. 3, for an image to be transmitted, we first obtain the patch sequence $\mathbf{P}$ after image tokenization, and we next process sequence $\mathbf{P}$ with a revised hyperprior variational (HV) vectorization network. The HV method, originally proposed in \cite{NTC2018}, can effectively model the spatial dependencies in the patch sequence $\mathbf{P}$ for improved image compression efficiency. We denote the latent representation sequence generated by the HV vectorization as $\mathbf{x} = [\mathbf{x}_1, \mathbf{x}_2, \ldots, \mathbf{x}_L]$, where $\mathbf{x}_i$ is a $c$-dimensional real-valued embedding vector and $c$ is the dimension hyperparameter of the embedding. In addition to the latent representation sequence $\mathbf{x}$, the HV vectorization also has a built-in entropy model that estimates the discrete entropy vector $\mathbf{e} = [e_1, e_2, \ldots, e_L]$ for given image patches according to the distribution parameter obtained from latent representations $\mathbf{x}$.

This paper revises the conventional HV training framework in \cite{NTC2018} by employing the MLLM as a teacher model and distilling its knowledge of SA semantic importance into the HV vectorization network serving as the student model. Specifically, the SA-semantic importance labels $\mathbf{I}$ from the MLLM are used as additional supervisory labels to guide the learning process of the student network. This revised training process allows the network to effectively learn the MLLM’s scenario-augmented knowledge, making entropy $\mathbf{e}$ characterize not only the structural/texture complexity as designed in \cite{NTC2018}, but also the scenario-relevant semantic importance of each patch, as emphasized in this paper. Henceforth, we refer to this revised entropy as \textit{SA entropy}.

Training details of the revised HV vectorization networks, as well as the re-designed loss function that facilitates the MLLM knowledge distillation, are available in Appendix \ref{appendixD}. Here in Section \ref{sec3}, we explain only the superiority of SA entropy over the conventional entropy from \cite{NTC2018} with the following example: in autonomous driving scenarios, a patch depicting the background may contain rich texture details such as distant buildings of different shapes, making the entropy of this patch relatively high when the conventional scenario-unaware HV network is considered. The revised HV networks distilled by the MLLM, however, understand that the shape of a background building, though rich in structural details, is irrelevant to the current driving scenario. Therefore, the SA entropy should be assigned to this background patch to facilitate later entropy-based patch compression in the JSCC encoding network.

With the above discussion, we write the HV vectorization process as:
$\{\mathbf{x}, \mathbf{e}\} = F_e(\mathbf{P})$, where $\mathbf{P}$ is the input patch sequence, $\mathbf{e}$ is the SA entropy vector, and $\mathbf{x}$ is the latent representation sequence vectorized from the patch sequence.

Subsequently, $\mathbf{x}$ and $\mathbf{e}$ are input into a variable-length JSCC encoder, which adaptively encodes $\mathbf{x}_i$ into a variable-length vector $\mathbf{y}_i$ with simultaneous consideration of both source coding and channel coding in conventional wireless communication systems. Ideally, as in previous investigations (e.g., \cite{NTSCC,NTSCC2}), the length of vector $\mathbf{y}_i$ reflects the coding resource allocated to image patch $i$, and therefore it should be proportional to $e_i$, the SA entropy of the patch.

However, the above ideal setup overlooks potential out-of-distribution (OOD) issues that may arise in real-world applications. OOD refers to situations where the model encounters samples or scenarios that fall outside the distribution of its training data \cite{OODProblem}. Such data may appear rarely or not at all during training but could occur occasionally in practical environments, bringing potential risk to the system if anti-OOD approaches are not considered. For example, autonomous vehicles may encounter novel traffic signs, construction machinery, or unusual obstacles arising from unexpected incidents. If the system fails to assign appropriate semantic importance to these OOD objects, its performance in real-world testing scenarios can be severely compromised. Recent studies have demonstrated that, owing to large-scale training datasets and diverse data sources, language models possess strong generalization capabilities in image analysis and are better equipped to handle the OOD problem \cite{addressOOD}. Thus, in addition to SA-entropy, our system further incorporates the SA-semantic importance labels assigned by the MLLM as an auxiliary preference to more accurately determine the length of the signal vector $\mathbf{y_i}$. This strategy is expected to combat the potential negative impact of OOD data on overall system performance.

Therefore, for each patch $i$, we jointly utilize its SA-entropy $e_i$ and SA-semantic importance label $I_i$ to determine $k_i$, the length of vector $\mathbf{y}_i$. Specifically, we let 
\begin{equation}
    k_i = Q_0 \left[ C_1(e_i) + C_2(I_i) \right], 
    \label{eq:vec_length}
\end{equation}
where $C_1(e_i)$ is the conventional vector length term that is proportional to the SA entropy (see (\ref{eq:c1}) in Appendix \ref{appendixF} for implementation details); $C_2(I_i)$ is the auxiliary vector length term developed to overcome the OOD problem with the MLLM (see (\ref{eq:c2}) in Appendix \ref{appendixF} for details); and $Q_0$ denotes a simple scalar quantizer whose duty is to map the sum $C_1(e_i) + C_2(I_i)$ into an integer within a pre-defined set of vector lengths supported by the communication system.\footnote{We refer readers to \cite{NTSCC} for theoretical justifications about the necessity of the scalar quantizer, and refer readers to (\ref{eq:quantizer}) in Appendix \ref{appendixF} for implementation details of the scalar quantizer in this paper.}

After obtaining $\mathbf{k} = [k_1, k_2, \ldots, k_L]$, we have, in essence, completed the dynamic encoding resource allocation with the help of the SA semantic importance. The rest of the task resembles the conventional JSCC methodology, i.e., coding image patch $i$ with the constraint of $k_i$ under the JSCC setup. Importantly, the MLLM’s scenario understanding is also involved in the training process by supervising the evaluation of the image distortion loss. In other words, the training of the JSCC encoder also follows the knowledge distillation setup discussed above. For details about the image distortion evaluation, we refer interested readers to Appendix \ref{appendixB}, while information about the loss function design is in Appendix \ref{appendixG}.

With the above discussion, we formally write the JSCC encoding process as $\mathbf{y} = G_e(\mathbf{e}, \mathbf{x}, \mathbf{I})$.
After the JSCC encoder, the complex-valued output $\mathbf{y}_i \in \mathbb{C}^{k_i}$. To satisfy the power constraint for signal transmission, vector $\mathbf{y} = [\mathbf{y}_1, \mathbf{y}_2, \ldots, \mathbf{y}_L]$ needs to undergo a power normalization process. The normalized channel input signal $\mathbf{y}^N$ satisfies the following constraint:
\begin{equation}
    \sum_{i=1}^L \| \mathbf{y}_i \|^2 \big/ {\sum_{i=1}^L k_i}   = 1,
    \label{eq:power_norm}
\end{equation}
where $\| \mathbf{y}_i \|^2$ denotes the squared norm of the complex-valued vector $y_i$, and $k_i$ is the encoding length for the corresponding patch. The normalized vector $\mathbf{y}^N = [\mathbf{y}_1^N, \mathbf{y}_2^N, \ldots, \mathbf{y}_L^N]$ is then transmitted over the noisy channel. Meanwhile, we note that the variable-length JSCC decoding process requires a priori knowledge about the vector length of each image patch. Therefore, we followed the setup in \cite{NTSCC} and transmitted $\mathbf{k} = [k_1, k_2, \ldots, k_L]$ through the channel.\footnote{We note that the a priori information of $k_i$ is vital for the JSCC decoding of patch $i$, just like the role of packet headers in ethernet network. To ensure reliable image reconstructions in the receiver, we followed the setup in \cite{NTSCC} and transmitted vector $\mathbf{k}$ with the communication scheme of low modulation order and high redundancy coding.}

\textbf{Wireless Channel:} When transmitted over the channel, the power-normalized signal $\mathbf{y}^N$ is passed through an additive white Gaussian noise (AWGN) channel, which is modeled as:
\begin{equation}
    \hat{\mathbf{y}} = H(\mathbf{y}^N) = \mathbf{y}^N + \mathbf{n},
    \label{eq:awgn}
\end{equation}
where $\mathbf{n} \sim \mathcal{CN}(0, \sigma_n^2)$ represents Gaussian noise with noise power $\sigma_n^2$, and $\hat{\mathbf{y}}$ denotes the received signal at the decoder.

\textbf{Receiver Design:} At the receiver side, the decoder aims to reconstruct the original image from the noise-corrupted received signal $\hat{\mathbf{y}}$, following a decoding process that mirrors the encoding procedure. Specifically, the received signal $\hat{\mathbf{y}}$ first passes through a variable-length JSCC decoder. Subsequently, the JSCC decoding output $\hat{\mathbf{x}} = [\hat{\mathbf{x}}_1, \hat{\mathbf{x}}_2, \ldots, \hat{\mathbf{x}}_L]$ is fed into the inverse HV vectorization module to generate the reconstructed patch sequence $\hat{\mathbf{P}} = [\hat{\mathbf{P}}_1, \hat{\mathbf{P}}_2, \ldots, \hat{\mathbf{P}}_L]$, which are then concatenated to form the reconstructed image $\hat{S}$.

We note that the JSCC decoder is jointly trained with the JSCC encoder, and the inverse HV vectorization network is also jointly trained with the HV vectorization network. That is, knowledge distillation is also applied in the training of receiver-side networks. For technical details about the joint network training, we refer readers to Appendix \ref{appendixH}.

\section{Model Training} \label{sec3b}
As illustrated in Fig. \ref{fig:sa-oosc-block}, this paper redesigns several key components within the conventional variable-length JSCC framework \cite{NTSCC, NTSCC2}. Due to space constraints, this section provides only an outline of our contributions in model training. We refer interested readers to appendices for technical details.

First, we present a revised evaluation scheme for image reconstruction distortion in Appendix \ref{appendixB}. This evaluation method emphasizes the reconstruction of scenario-relevant information through a SA weighted mechanism, serving as a critical loss term in the network training process. Subsequently, Appendix \ref{appendixC} presents the architectural improvements made to the conventional HV vectorization and inverse vectorization networks, while Appendix \ref{appendixD} details their loss function design. And this is followed by the network implementation details in Appendix \ref{appendixE}. Meanwhile, variable-length JSCC encoder and decoder are trained, with their implementation details and loss functions introduced in Appendices \ref{appendixF} and \ref{appendixG}, respectively. The JSCC encoder addresses the OOD problem identified in \cite{addressOOD} through an auxiliary loss term based on the MLLM's importance identification. Finally, Appendix \ref{appendixH} presents the overall system training framework and hyperparameter configuration.

\section{Experiments} \label{sec4} 
\begin{figure*}[htbp]
    \centering
    \includegraphics[width=1.0\linewidth]{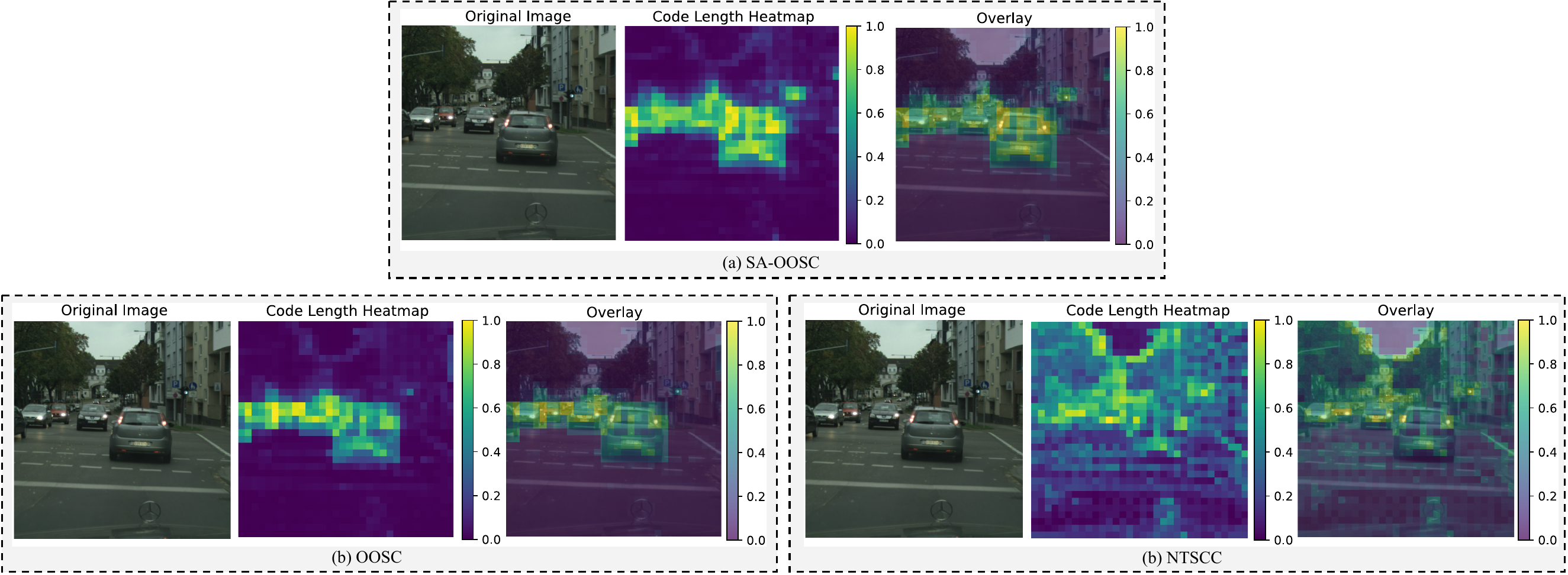} 
    \captionsetup{font={small}}
    \caption{Comparison of code rate distributions under different schemes. For each scheme, the left column shows the original image, the middle column presents the code length heatmap, and the right column displays the overlay of the image and its code length distribution. The heatmap reflects the allocation of channel resources across image patches, with brighter colors indicating a larger number of channel symbols assigned. The overlay intuitively visualizes the spatial regions prioritized by each encoder. To ensure a fair comparison, all schemes are configured to have similar CBR (0.0228, 0.0230 and 0.0241 for SA-OOSC, OOSC, and NTSCC, respectively).}
    \label{fig:code_distribution}
\end{figure*}

This paper has comprehensively evaluated the two major technical contributions of this paper: 1) the SA-semantic importance identification empowered by MLLM, and 2) the SA-OOSC framework. Due to page limit, we cannot present all experimental details here. Discussion about the MLLM-empowered SA semantic importance identification are given in Appendix \ref{sec4sub1}, while the rest of this section details experiment and associated results for testing the SA-OOSC framework. Experimental setups, including dataset applied, baseline methods, and evaluation metrics, are given as follows.\footnote{For the hyperparameter setup of our model tested in the experimental section, we refer readers to Appendix \ref{appendixH} of the paper. Regarding the implementation of baseline methods, we refer readers to our baseline implementation open-sourced along with the data benchmark.}

\textbf{Dataset:} We utilized the Cityscapes dataset \cite{Cityscapes} to evaluate the SA-OOSC framework, using the autonomous driving scenarios as a proof-of-concept. The dataset consists of 2975 training images and 1525 testing images, each with a resolution of $2048 \times 1024$ pixels.

\textbf{Baseline Methods Benchmarked:} We compared our method with the representative baselines as follows: 1) the fixed-rate image transmission scheme (Deep JSCC) proposed in \cite{Deepjscc}, 2) the adaptive-rate transmission scheme (NTSCC) developed in \cite{NTSCC}, and 3) the OOSC scheme introduced in \cite{OOSC}.

\textbf{Evaluation Metrics:} Following \cite{Deepjscc}, Channel Bandwidth Ratio (CBR) was considered as the main metric for evaluating the compression efficiency of image transmission. The original bandwidth was defined as $n = 3hw$, where $h$ and $w$ denoted the height and width of the image, respectively. Considering an image with $L$ patches, the encoded image could be represented by $L$ vectors of varying lengths, with the $i$-th vector having length $k_i$. The total bandwidth cost was $m = \sum_{i=1}^L k_i$, and the system’s CBR was defined as $R = m/n$ to represent the average number of available channel symbols per source dimension. PSNR was used to assess the quality of reconstructed image patches.

We start with a case study at the transmitter side. Fig.~\ref{fig:code_distribution} compares our method with OOSC and NTSCC. The fixed-rate Deep JSCC approach is not included here in this case study, as it does not support semantic-driven adaptive rate allocation. In this experiment, an image was randomly selected from the dataset to illustrate the normalized code rate distributions under different encoding strategies.

Visualizations in Fig. \ref{fig:code_distribution} demonstrate that our method could more precisely allocate channel resources to patches with high SA semantic importance, effectively reducing bandwidth assigned to less relevant background areas. For example, in this urban driving scene, the vehicle immediately in front of the ego car should be assigned the highest semantic importance, while other vehicles farther away, although still relevant, should have lower priority. SA-OOSC adaptively captured this semantic hierarchy and reflected such priority in channel resource allocation. The OOSC scheme, on the other hand, allocated resources rather coarsely based on predefined object categories (e.g., vehicles, pedestrians), lacking the ability to distinguish importance among similar objects in different spatial contexts. Hence, it failed to assign sufficient resources to the most critical vehicle (i.e., the closest car), while expending unnecessary bandwidth on less relevant vehicles in other lanes. The NTSCC scheme, while capable of adaptive rate allocation, did not explicitly incorporate task-relevant semantic importance, distributing resources primarily according to image texture and detail -- substantial resources were wasted on irrelevant regions such as background buildings.\footnote{Besides the transmitter-side case study presented, we also have a case study that took the noisy channel and image reconstruction into account. We refer readers to Appendix \ref{appendixI} for details about this receiver-side case study.}

\begin{figure}[htbp]
    \centering
    \includegraphics[width=0.9\linewidth]{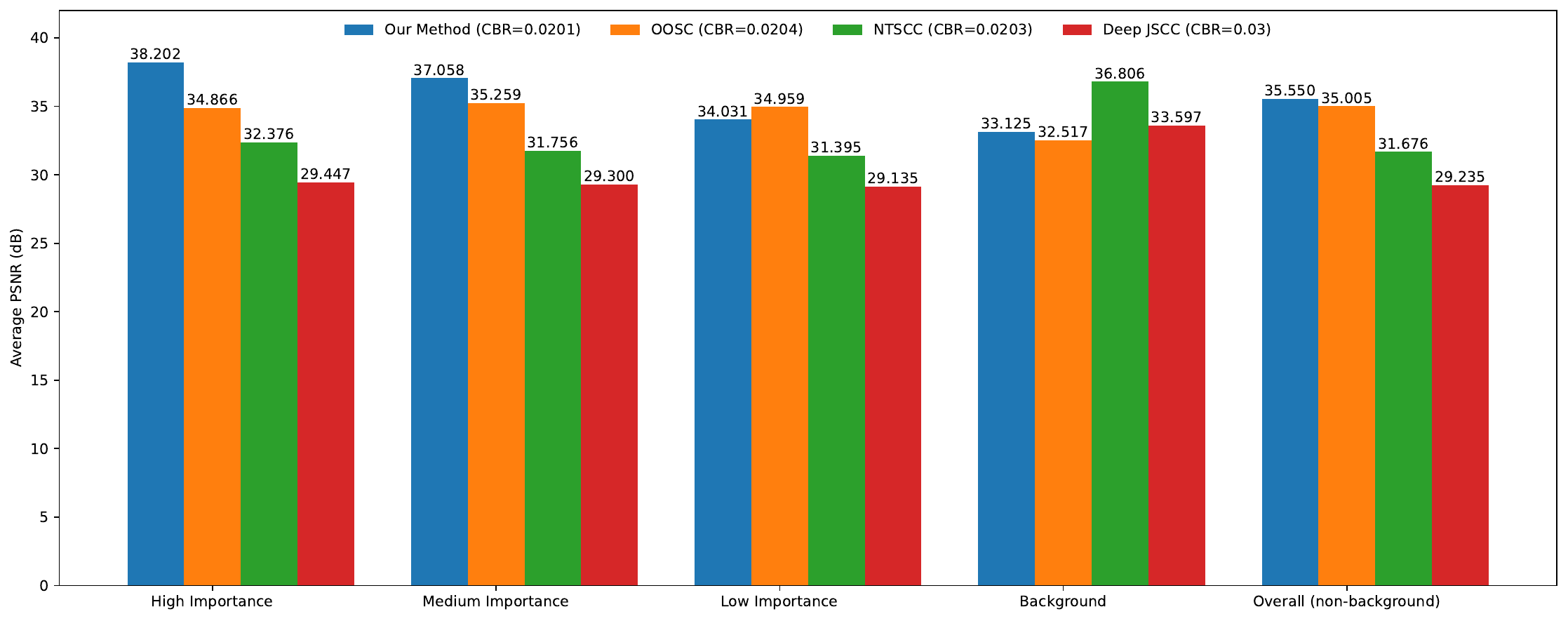} 
    \captionsetup{font={small}}
    \caption{Image reconstruction evaluation for batches with different semantic importance.}
    \label{fig:psnr_semantic}
\end{figure}

\begin{figure*}[t]
    \centering
    \includegraphics[width=1.0\linewidth]{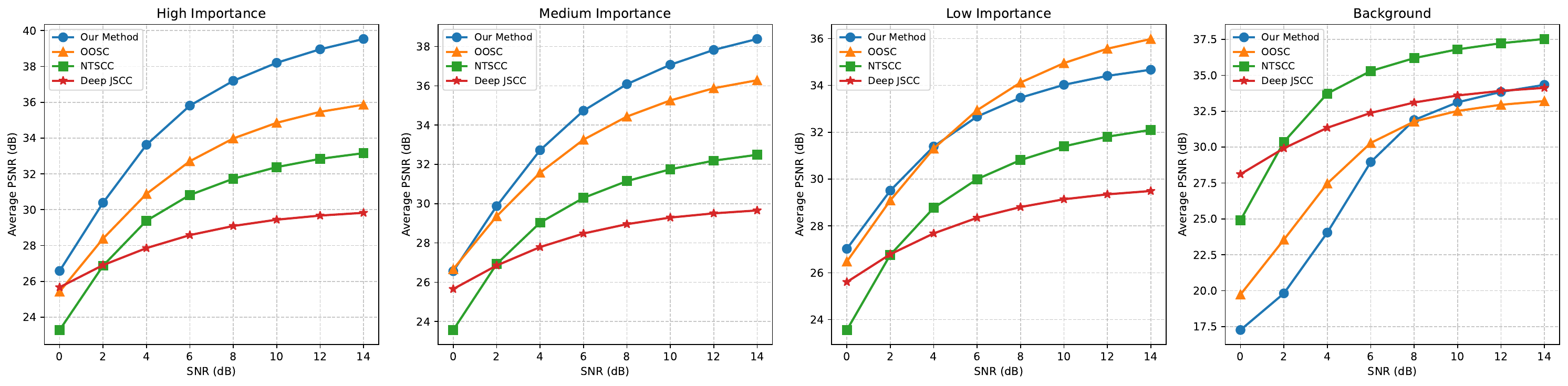}
    \captionsetup{font={small}}
    \caption{Image reconstruction across semantic importance levels under varying SNR conditions.}
    \label{fig:psnr_snr}
\end{figure*}

Experiments in Fig. \ref{fig:psnr_semantic} further evaluated the performance of the proposed SA-OOSC method with the whole dataset, in which PSNR was compared across image patches with different levels of semantic importance. Both the training and the testing phase assumed 10~dB SNR. As the figure shows, SA-OOSC achieved superior overall channel bandwidth efficiency, with its 0.0201 CBR lowering than other baseline methods. Despite the low channel bandwidth occupation, SA-OOSC consistently achieved the best reconstruction performance in both high and medium semantic importance patches. In low-importance patches, OOSC slightly outperformed SA-OOSC, while in background patches, both SA-OOSC and OOSC exhibited low PSNR, as these methods intentionally allocated fewer resources to background content. NTSCC and Deep JSCC achieved higher PSNR in the background, although the information therein was less important. From an overall (non-background) perspective, SA-OOSC achieved an average PSNR of 35.55~dB, which was comparable to OOSC (35.01~dB) but significantly higher than NTSCC (31.68~dB) and Deep JSCC (29.24~dB). These results clearly demonstrate that SA-OOSC could prioritize the reconstruction of patches with scenario-relevant information, maintaining a high image quality with the lowest CBR.

Fig. \ref{fig:psnr_snr} benchmarks SA-OOSC with baseline methods under various SNR. Here we analyzed the average PSNR of different methods across different patches as SNR varies, with all models trained with 10~dB SNR. For patches with high semantic importance, SA-OOSC consistently achieved the best PSNR. In medium-importance patches, SA-OOSC maintained the highest PSNR across all SNR settings, demonstrating strong noise robustness in critical target areas. For low-importance patches, OOSC and SA-OOSC performed similarly, both outperforming NTSCC and Deep JSCC. For background patches, our method did not perform as well as NTSCC and Deep JSCC. Yet, due to limited impacts of background patches on decision-making, the minor degradation in reconstruction quality is acceptable.

\section{Conclusion} \label{sec5} 
This paper proposes a MLLM-empowered semantic identification method with task scenario understanding. Meanwhile, we develop SA-OOSC, an image semantic communication framework that leverages the MLLM-based identification for knowledge distillation, enabling the framework to understand task-relevant scenarios. Experiments show that SA-OOSC achieves superior coding efficiency while preserving all important scenario-relevant information, demonstrating robust image reconstruction across various channel conditions. Further, we open-source the MLLM-annotated dataset with manual sampling and cross-checking, which represents an vital data
contribution to the research community of semantic communication.

\bibliographystyle{IEEEtran}
\bibliography{Main}

\appendices

\section{Implementation Details of SA-Semantic Importance Labeling Module} \label{appendixA}
The pipeline of the SA-semantic importance labeling module has been detailed in section \ref{sec2}. In this section, we further elaborate on the specific implementation process, which consists of two main stages: object detection and importance assignment. In the object detection stage, we utilize the high-performance object detector YOLOv11x \cite{yolov11x} to identify objects relevant to the traffic scenario. This model demonstrates a high recall rate for key traffic objects. In the importance assignment stage, we employ GPT-4 \cite{gpt4} as a multi-modal annotator to assign SA-semantic importance levels to the detected objects within the traffic scenario. The prompt is carefully designed to simulate the driver’s perspective, guiding the model to comprehensively consider each object’s distance from the ego vehicle, its relevance to the driving path, and its potential impact on driving decisions. The prompt incorporates rule-based annotation criteria while also granting the model the flexibility to perform contextual reasoning in complex scenarios.

The prompt is designed through a multi-step process. We first instruct the model to assess each object in the context of the traffic scenario, considering its distance from the ego vehicle and its potential influence on driving behavior. Based on these spatial and semantic factors, the model is expected to assign \emph{low}, \emph{medium}, or \emph{high} to each object. We observed that certain objects (e.g., traffic lights) exhibited higher uncertainty due to ambiguous visual context. To address this, we incorporated rule-based constraints into the prompt to explicitly guide the model to assign higher importance to such critical traffic elements, thereby improving consistency and alignment with real-world driving priorities. However, the model occasionally assigns low importance to objects at a close distance, due to their appearance as parked vehicles or non-standard traffic participants. To mitigate this, we enforce a rule that all nearby objects should be labeled as ``high'' regardless of the motion status. Additionally, we include examples in the prompt to provide specific guidance and improve the model’s accuracy in handling such cases.

The detailed prompt is as follows:
\begin{lstlisting} 
You are a strict and objective evaluator, your task is to evaluate the importance of objects in the image based on the images, Detection Information and Evaluation Criteria. 

Given the two images, the first one is the original image and the second one is image with detected objects with their ids. The detection information is a json string comes, each component is formatted as:
{'name': <name of the object>, 'class': <integer representing the object>, 'confidence': <double number>, 'box': {'x1': <double number>, 'y1': <double number>, 'x2': <double number>, 'y2': <double number>}, 'track_id': <id of the object>}, where x1y1 is the top-left corner of the bounding box, and x2y2 is the bottom-right corner of the bounding box. [Detection Information]:\n

YOLO detection information [Evaluation Criteria]: There are three level of importance: 1, 2, 3. 1 for lowest importance, 2 for middle importance, 3 for highest importance. Based on the detection information, assume that in a driving scenery, you are the driver seeing this scene, in which the car at the bottom of the image is the car that you are driving, the car you are driving itself is not important. You should consider the relative position of other objects to the car you are driving. Objects that can influence the driving situation should be given higher importance, while objects that are not relevant to the driving situation or cannot influence the driving situation should be given lower importance although they are close to the car. Objects (e.g., cars, people) in the same path should be given higher importance, while objects that are not in the path (e.g., driving in another road with opposite direction) should be given lower importance. Objects that are very close to the car you are driving (e.g. cars) should be given higher importance. You should compare the distance of each object to the car you are driving, the closest ones should be given 3, and the farthest ones should be given 1. 

Specifically, if objects are very close to you, you should allocate 3 to them, even if they are not moving (for example, a parked car which is very close to you). Also, you should pay attention to the traffic signs and signals, and those are relevant to the driving situation should be given higher importance. For example, those traffic lights directly guiding the car you are driving should be allocated 3. During the evaluation, you should consider every object in the image, and think step by step. Allocate level of importance to each objects in the image. The evaluation and your output must be strictly structured in the following JSON format without any comments:
{<object_id>: <importance_level>,......,<object_id>: <importance_level>}
\end{lstlisting}

With the above prompt, the MLLM assigns an object-based SA-semantic importance label to each detected object. As described in section \ref{sec2}, these object-based labels are subsequently converted to the corresponding patch-based SA-semantic importance labels. This data contains scenario understanding knowledge from the MLLM and is used for knowledge distillation to equip the entire framework with scenario understanding capabilities.

\section{Weighted Evaluation for the Image Reconstruction Distortion} \label{appendixB}

This section explains the evaluation metric proposed in this work, which effectively quantifies the comprehensive reconstruction quality of the image by accounting for the scenario-augmented semantic importance of different regions.

Unlike traditional semantic communication systems that evaluate the communication quality with the distortion between the reconstructed image and the original one solely, such as those previous research that relied only on the peak signal-to-noise ratio (PSNR) for evaluation, this paper places greater emphasis on the task-relevant semantic importance. We aim to achieve better reconstruction quality for image regions with higher semantic importance in the given context, thereby enhancing task relevance. To this end, we propose to use a SA-semantic importance weighted distortion metric for evaluating the quality of scenario-aware image reconstruction.

To facilitate importance weighted distortion, we map importance label $I_i$ of image patch $i$ to a weight $w_i$, which reflects its contribution of the patch to the overall distortion. The weights are defined using an exponentially weighted normalization formula:
\begin{equation}
w_i = \frac{2^{I_i}}{\sum_{j=1}^L 2^{I_j}},
\label{eq:weight}
\end{equation}
where $I_i$ is set as $3, 2, 1, 0$ for high, medium, low, and background semantic importance, respectively. For patch $i$, we follow the classic definition of PSNR and compute its patch-wise distortion metric as
\begin{equation}
\mathrm{PSNR}_i = 10 \log_{10} \left( \frac{\mathrm{MAX}^2}{\frac{1}{M} \sum_{m=1}^M (P_i[m] - \hat{P}_i[m])^2} \right),
\label{eq:psnr_patch}
\end{equation}
where $P_i[m]$ and $\hat{P}_i[m]$ represent the $m$-th pixel value in the original and reconstructed image patch, respectively; $\mathrm{MAX}$ is the maximum possible pixel value (e.g., $\mathrm{MAX}=255$ for 8-bit images), and $M$ is the total number of pixels in each patch.

The overall distortion metric for the image is then obtained by computing the weighted average of the PSNR values of all patches:
\begin{equation}
\mathrm{SAD} = \sum_{i=1}^L w_i \cdot \mathrm{PSNR}_i,
\label{eq:sad}
\end{equation}
where $\mathrm{SAD}$ represents the SA-semantic importance weighted image distortion, or SA-Distortion in short.

\section{Architectural Designs of HV Vectorization and Inverse Vectorization Networks}
\label{appendixC}

This section provides a detailed exposition of the architecture design of the HV vectorization and inverse vectorization modules in our paper, including the processes by which an image $S$ is vectored into its latent representation $\mathbf{x}$ and the corresponding SA-entropy $\mathbf{e}$, as well as how $\mathbf{x}$ is decoded to reconstruct the image $\hat{S}$. As the architecture of the modules follows that of the hyperprior variational image compression framework proposed in \cite{NTC2018} (to be replaced with your actual citation), readers are referred to \cite{NTC2018} for further details.

\begin{figure*}[htbp]
    \centering
    \includegraphics[width=0.75\linewidth]{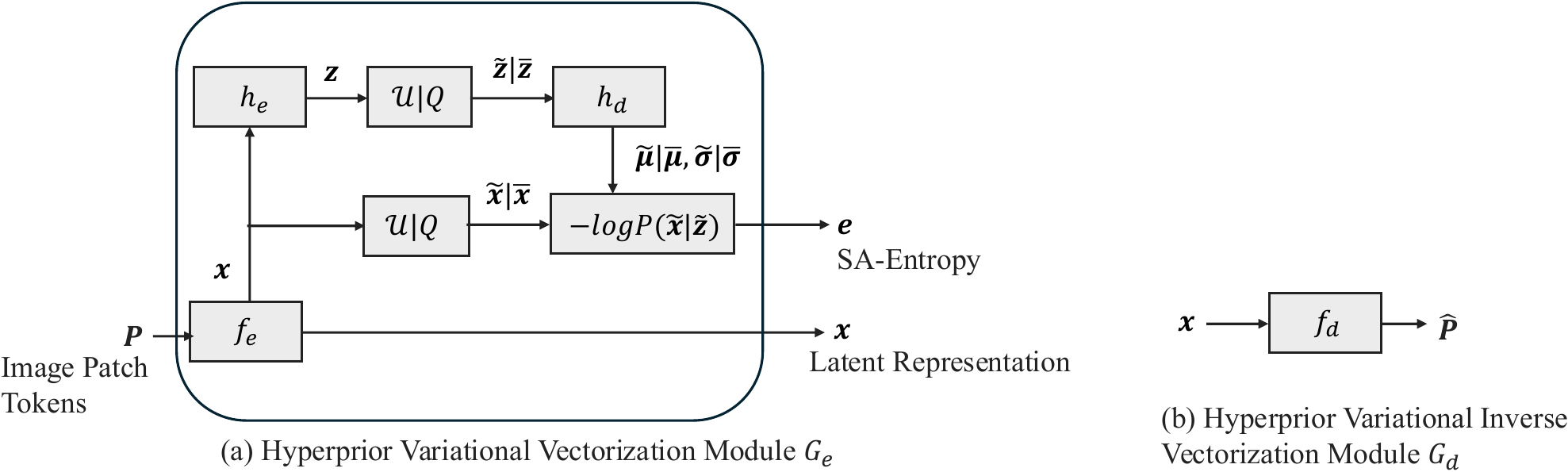}
    \captionsetup{font={small}}
    \caption{The architecture of the HV Vectorization (a) and Inverse Vectorization (b) Modules.}
    \label{fig:hv_architecture}
\end{figure*}

As introduced in section \ref{sec2}, the patch token sequence $\mathbf{P}$ is mapped by the HV vectorization into a corresponding latent representation vector $\mathbf{x} = [\mathbf{x}_1, \ldots, \mathbf{x}_L]$, along with the associated SA-entropy sequence $\mathbf{e} = [e_1, \ldots, e_L]$. For simplicity in the subsequent theoretical derivations, we treat the image as a single patch, and the following analysis is conducted on this unified patch. Under this assumption, $\mathbf{P}$ is equivalent to the original image $S$. As the processing of the patch sequence essentially involves parallel operations on individual patches, this simplification does not affect the generality or validity of the subsequent analysis. The internal structure of the HV vectorization is illustrated in Fig. \ref{fig:hv_architecture}a, which serves as the basis for the following detailed explanation of the encoding process.

The HV vectorization is a lossy compression model that simulates the classical source coding process. The key idea is that the source image $S$ is not directly mapped to a codeword vector, but is first passed through a variational encoder $f_e$ parameterized by neural network parameters $\theta_f$, which implements a nonlinear transformation from the original image space to the latent space, yielding the latent representation $\mathbf{x} = f_e(S; \theta_f)$. The latent representation $\mathbf{x}$ preserves the semantic features of the source signal, and its dimension is typically much smaller than the original dimension. In the latent space, $\mathbf{x}$ is modeled by a conditional Gaussian distribution, whose prior is determined by an additional latent variable $\mathbf{z}$.

\textbf{Remark:} It is worth noting that, in traditional autoencoder-based compression methods, it is typically assumed that the components of the latent vector $\mathbf{x}$ are independent (i.e., they follow a fully factorized prior \cite{NTC2017}). However, Ballé et al. \cite{NTC2018} observed that the learned latent representations $\mathbf{x}$ are often highly correlated in practice, especially in edge and texture regions. This leads to clustering of latent values, making the independence assumption inadequate for modeling the true distribution and thus limiting entropy coding efficiency. To address this, Ballé et al. introduced an auxiliary latent variable $\mathbf{z}$ to capture dependencies among components of $\mathbf{x}$, using a transformation similar to that from $\mathbf{P}$ to $\mathbf{x}$. This approach enables more effective learning of the structural features of images for compression.

Specifically, a hyperprior encoder $h_e$ is cascaded after the variational encoder $f_e$ to further extract contextual information from $\mathbf{x}$, resulting in $\mathbf{z} = h_e(\mathbf{x}; \theta_h)$, where $\theta_h$ denotes its parameters. As we discussed in section \ref{sec3}, the SA entropy of the latent representation $\mathbf{x}$ produced by the HV vectorization module is further utilized in this work to guide the rate allocation in the variable-length JSCC encoder. For computational tractability, we use discrete entropy rather than differential entropy to estimate the coding rate required for $\mathbf{x}$ and $\mathbf{z}$, which necessitates quantization. This is achieved by rounding each component of $\mathbf{z}$, resulting in $\bar{\mathbf{z}}$. During training, since quantization is non-differentiable, Ballé et al. proposed to approximate this process by adding uniform noise $\mathbf{o} \sim \mathcal{U}(-0.5, 0.5)$ to $\mathbf{z}$, i.e., $\tilde{\mathbf{z}} = \mathbf{z} + \mathbf{o}$, thereby allowing gradients to propagate \cite{NTC2018}; actual quantization is used during inference.

The distribution of $\tilde{\mathbf{z}}$ is modeled as a fully factorized non-parametric density \cite{NTC2017}. A fully factorized density assumes that the components of the vector are statistically independent, so the joint distribution can be decomposed as the product of the marginal distributions of each component. The "non-parametric" aspect means that the form of each marginal distribution is not restricted to a specific parametric family (such as Gaussian) but is instead learned flexibly. Formally, the density is expressed as
\begin{equation}
    p_{\tilde{\mathbf{z}}|\psi} (\tilde{\mathbf{z}}|\psi) = \prod_{j} \left[ p_{z_j|\psi^{(j)}}(z_j|\psi^{(j)}) * \mathcal{U}(-\frac{1}{2}, \frac{1}{2}) \right](\tilde{z}_j)
    \label{eq:hyperprior_density}
\end{equation}
where $p_{z_j|\psi^{(j)}}(z_j|\psi^{(j)})$ is the prior for each component, $j$ indexes the components of $\tilde{\mathbf{z}}$, $\psi$ collectively represents the set of parameters that define the prior distributions for all components, $\psi^{(j)}$ is the subset of parameters specifically associated with the $j$-th component, and $*$ denotes the convolution operation.

The quantized or noise-perturbed $\mathbf{z}$ is then passed through the hyperprior decoder $h_d$ to predict the parameters of the Gaussian distribution $(\tilde{\boldsymbol{\mu}}, \tilde{\boldsymbol{\sigma}}) = h_d(\tilde{\mathbf{z}}; \phi_h)$, where $\phi_h$ are the parameters of the hyperprior decoder. Finally, the quantized $\mathbf{x}$ and $\mathbf{z}$ are input to the entropy model to obtain the corresponding discrete entropy, which guides the subsequent allocation of channel transmission rate. The entropy model is computed via the conditional probability $p_{\tilde{\mathbf{x}}|\tilde{\mathbf{z}}}$:
\begin{equation}
    p_{\tilde{\mathbf{x}}|\tilde{\mathbf{z}}} (\tilde{\mathbf{x}}|\tilde{\mathbf{z}}) = \prod_i \left[ \mathcal{N}(\tilde{\mu}_i, \tilde{\sigma}_i^2) * \mathcal{U}(-\tfrac{1}{2}, \tfrac{1}{2}) \right](\tilde{x}_i)
    \label{eq:entropy_model}
\end{equation}
where $i$ indexes the elements of the latent vector $\tilde{\mathbf{x}}$, and $\mathcal{N}(\tilde{\mu}_i, \tilde{\sigma}_i^2)$ denotes a Gaussian distribution with mean $\tilde{\mu}_i$ and variance $\tilde{\sigma}_i^2$ for the $i$-th component $\tilde{x}_i$.

The discrete entropy of the latent representation is then given by
\begin{equation}
    \mathbf{e} = -\log P(\tilde{\mathbf{x}}|\tilde{\mathbf{z}})
    \label{eq:discrete_entropy}
\end{equation}
Through the above process, the encoder outputs both the latent representation $\mathbf{x}$ and the corresponding discrete entropy $\mathbf{e}$ for the image $S$. As shown in Fig. \ref{fig:hv_architecture}b, during the decoding process, the quantized latent representation $\tilde{\mathbf{x}}$ is fed into the variational decoder $f_d$ to obtain the reconstructed image $\hat{S} = f_d(\tilde{\mathbf{x}}; \phi_f)$.

\section{Loss Functions for Training HV Vectorization and Inverse Vectorization Networks}
\label{appendixD}

The optimization problem for the HV vectorization module can essentially be modeled as a variational autoencoder (VAE) \cite{vae}. The objective of this module is to construct a parameterized variational density $q(\tilde{\mathbf{x}}, \tilde{\mathbf{z}}|S)$ to approximate the true but intractable posterior $p(\tilde{\mathbf{x}}, \tilde{\mathbf{z}}|S)$, thereby enabling efficient generative modeling and compression. Specifically, under the source distribution $p_S$, the model minimizes the Kullback-Leibler (KL) divergence between the approximate and true posterior distributions, which can be formalized as:
\begin{equation}
\min_{\theta_f,\, \theta_h,\, \phi_g,\, \phi_h} \;\;
\mathbb{E}_{S \sim p_S} \left[
\mathrm{KL}\left(q(\tilde{\mathbf{x}}, \tilde{\mathbf{z}}|S)\;\|\;p(\tilde{\mathbf{x}}, \tilde{\mathbf{z}}|S)\right)
\right].
\label{eq:hv_kl}
\end{equation}

Since the true posterior $p(\tilde{\mathbf{x}}, \tilde{\mathbf{z}}|S)$ is typically intractable, the above optimization problem is equivalent to maximizing the evidence lower bound (ELBO) of the observed data log-likelihood~\cite{vae}:
\begin{equation}
\max \;\; \mathbb{E}_{S \sim p_S} \; \mathbb{E}_{\tilde{\mathbf{x}}, \tilde{\mathbf{z}} \sim q(\tilde{\mathbf{x}}, \tilde{\mathbf{z}}|S)} \left[
\log
\frac{p_{\tilde{\mathbf{z}}}(\tilde{\mathbf{z}})
p(\tilde{\mathbf{x}}|\tilde{\mathbf{z}})
p(\hat{S}|\tilde{\mathbf{x}})}
     {q(\tilde{\mathbf{x}}, \tilde{\mathbf{z}}|S)}
\right],
\label{eq:hv_elbo}
\end{equation}
where $p_{\tilde{\mathbf{z}}}(\tilde{\mathbf{z}})$ denotes the prior distribution of the hyperprior latent variable, $p(\tilde{\mathbf{x}}|\tilde{\mathbf{z}})$ is the conditional distribution of the primary latent code, $p(\hat{S}|\tilde{\mathbf{x}})$ is the likelihood function for the reconstructed image, and $q(\tilde{\mathbf{x}}, \tilde{\mathbf{z}}|S)$ is the approximate posterior along the encoding path.

During training, due to the use of additive uniform noise to simulate quantization, $q(\tilde{\mathbf{x}}, \tilde{\mathbf{z}}|S)$ can be regarded as a constant distribution, whose log-probability becomes a constant term and can be omitted during parameter updates. Consequently, the final loss function simplifies to the expected negative log-likelihood of the numerator:
\begin{equation}
\mathcal{L} = \mathbb{E}_{S \sim p_S} \left[
    -\log p_{\tilde{\mathbf{z}}}(\tilde{\mathbf{z}})
    -\log p(\tilde{\mathbf{x}}|\tilde{\mathbf{z}})
    -\log p(\hat{S}\,|\,\tilde{\mathbf{x}})
\right],
\label{eq:hv_loss_basic}
\end{equation}
where $-\log p_{\tilde{\mathbf{z}}}(\tilde{\mathbf{z}})$ measures the transmission rate consumption for encoding the hyperprior side information; $-\log p(\tilde{\mathbf{x}}|\tilde{\mathbf{z}})$ quantifies the transmission rate for the primary latent representation $\tilde{\mathbf{x}}$; and $-\log p(\hat{S}\,|\,\tilde{\mathbf{x}})$ reflects the negative log-likelihood of the reconstructed image.

Given that this work focuses on image compression guided by SA semantic importance, the third term, $-\log p(\hat{S}\,|\,\tilde{\mathbf{x}})$, is instantiated as the SA Semantic Importance-weighted Distortion (SAD) function (see Appendix \ref{appendixB} for calculation details). To enable flexible trade-offs between rate and distortion, a weighting parameter $\lambda$ is introduced. The joint loss function $\mathcal{L}_{\mathrm{hv}}$ for the HV vectorization and inverse vectorization modules is formulated as follows:
\begin{equation}
\mathcal{L}_{\mathrm{hv}} =
\mathbb{E}_{S \sim p_S} \left[
    \lambda \left(
        -\log p(\tilde{\mathbf{x}}|\tilde{\mathbf{z}})
        -\log p_{\tilde{\mathbf{z}}}(\tilde{\mathbf{z}})
    \right)
    + \mathrm{SAD}(S, \hat{S}_{\mathrm{hv}})
\right],
\label{eq:hv_final_loss}
\end{equation}
where $\mathrm{SAD}(S, \hat{S}_{\mathrm{hv}}) = \sum_{i=1}^L w_i \cdot \mathrm{PSNR}_i$, $\hat{S}_{\mathrm{hv}}$ denotes the reconstructed image obtained by directly decoding the source image $S$ using only the HV vectorization and inverse vectorization modules, $w_i$ is the semantic importance weight for the $i$-th patch, and $\mathrm{PSNR}_i$ is the PSNR of the corresponding patch reconstruction.

It is important to note that after the introduction of the SAD, the mathematical forms of the probability modeling terms in the loss function (i.e., $-\log p(\tilde{\mathbf{x}}|\tilde{\mathbf{z}})$ and $-\log p_{\tilde{\mathbf{z}}}(\tilde{\mathbf{z}})$) remain unchanged, corresponding to the theoretically optimal transmission rates for the primary latent representation and the hyperprior side information, respectively. However, as the model assigns greater reconstruction weight to semantically important regions during optimization, the network parameters adaptively adjust the distribution of latent representations to accommodate the corresponding semantic preferences. This enables the HV network to learn the knowledge of SA semantic importance from the MLLM. Such a knowledge distillation approach aligns the scenario understanding capability of the HV network with that of the MLLM, thereby allowing for more accurate estimation of the SA entropy for each image patch.

\section{Implementation Details of HV Vectorization and Inverse Vectorization Networks}
\label{appendixE}

\begin{figure*}[htbp]
    \centering
    \includegraphics[width=0.95\linewidth]{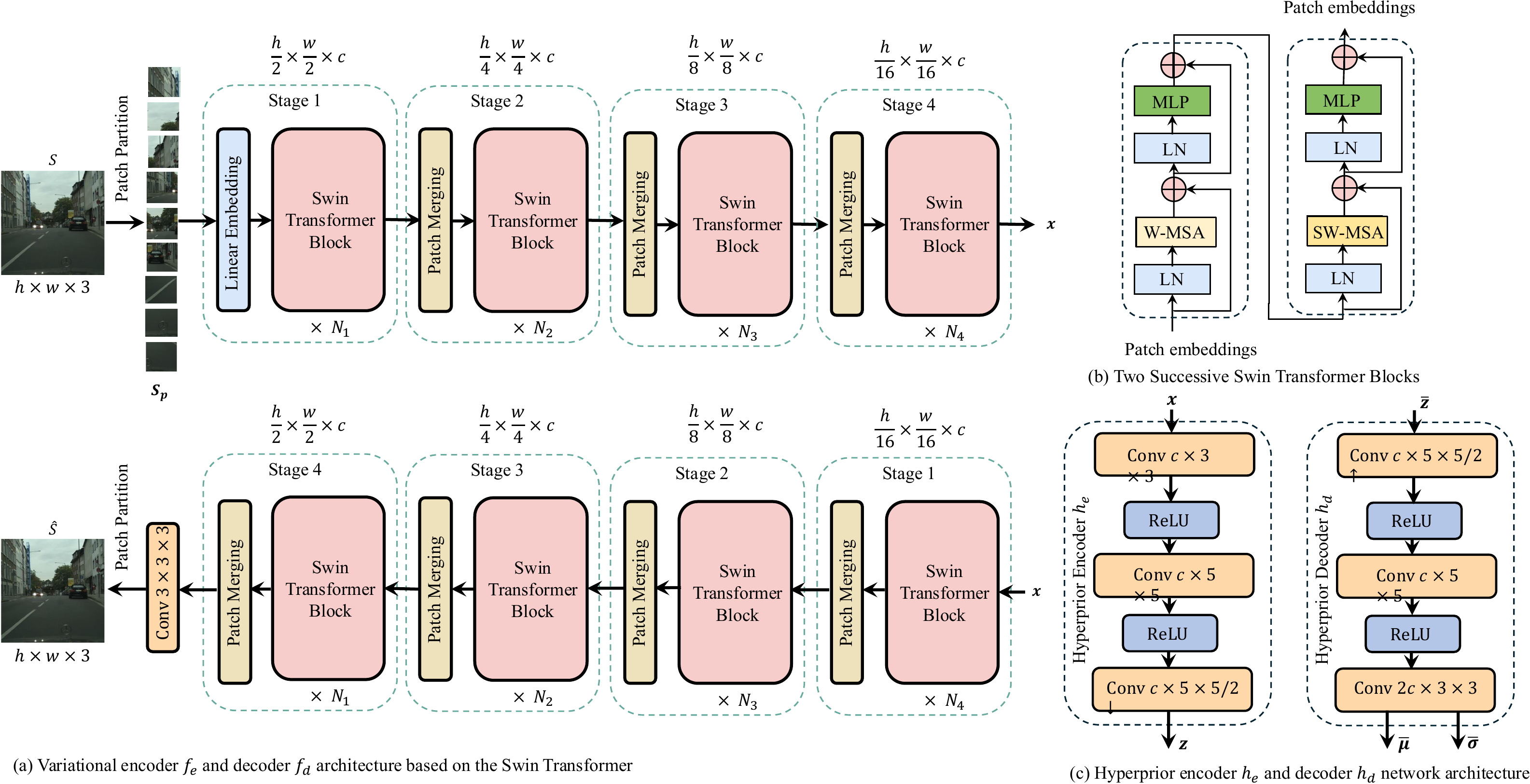} 
    \captionsetup{font={small}}
    \caption{The implementation of components involved in HV vectorization and inverse vectorization modules. (a) Overall architecture based on Swin Transformer; (b) The detailed structure of a Swin Transformer Block; (c) Architecture of the hyperprior encoder and decoder.}
    \label{fig:swin_arch}
\end{figure*}

The neural network architecture adopted in the HV modules consists of four nonlinear transformation modules: the variational encoder $f_e$, variational decoder $f_d$, hyperprior encoder $h_e$, and hyperprior decoder $h_d$. Specifically, the input RGB image $S \in \mathbb{R}^{h \times w \times 3}$ is first uniformly partitioned into $l_1 = h/2 \times w/2$ non-overlapping patches, each with a size of $2 \times 2 \times 3$, forming a patch token sequence $\mathbf{P}$. As shown in Fig. \ref{fig:swin_arch}a, the architecture of the variational encoder and decoder follows the Swin Transformer framework proposed in \cite{swintransformer}. During variational encoding, each patch token is first projected via a linear embedding layer to a $c$-dimensional feature vector. The embedded patch tokens are then input into $N_1$ stacked Swin Transformer Blocks for feature extraction.

Fig. \ref{fig:swin_arch}b illustrates the detailed structure of a Swin Transformer Block \cite{swintransformer}. Each block consists of two sequential window-based attention modules: Window-based Multi-head Self-Attention (W-MSA) and Shifted Window Multi-head Self-Attention (SW-MSA), which alternately process the features. In each module, the input patch embeddings are first normalized by LayerNorm and then passed through the multi-head self-attention mechanism. The W-MSA module partitions the input features into several non-overlapping local windows and computes multi-head self-attention independently within each window, effectively modeling spatial correlations in local regions and capturing fine-grained details with low computational complexity. The SW-MSA module introduces a window shifting operation -- shifting the window partitions by half the window size -- enabling information interaction across windows, thus enhancing global feature modeling. Residual connections are employed after each attention sub-layer to facilitate gradient flow and information preservation. The output of the attention layer is then normalized again and passed through a two-layer feed-forward MLP, which incorporates nonlinear activations (such as GELU) to improve representation capacity. The MLP output is also combined with the previous features via residual connection to produce the final block output.

At the end of each stage, a \emph{Patch Merging} operation concatenates the embeddings of every $2 \times 2$ group of neighboring patches into a $4c$-dimensional feature, which is then linearly projected back to $c$ dimensions, thus achieving a twofold reduction in spatial resolution. The resulting feature sequence (of length $l_2 = h/4 \times w/4$) is then fed into the next set of Swin Transformer Blocks. The entire variational encoder comprises four such stages, each containing several Swin Transformer Blocks and a Patch Merging operation, ultimately producing the main latent variable $\mathbf{x} \in \mathbb{R}^{h/16 \times w/16 \times c}$. The variational decoder $f_d$ adopts a symmetric architecture to $f_e$, utilizing the inverse of Patch Merging (i.e., Patch Division) and progressive upsampling, in conjunction with Swin Transformer Blocks, to gradually restore the spatial resolution and ultimately reconstruct the output image $\hat{S}$.

The hyperprior autoencoder, as depicted in Fig.~\ref{fig:swin_arch}c, comprises the hyperprior encoder $h_e$ and hyperprior decoder $h_d$, which are mainly responsible for modeling the distribution parameters of the latent variables and assisting in entropy coding. The hyperprior encoder $h_e$ takes the main latent variable $\mathbf{x}$ as input and consists of multiple convolutional layers (Conv + ReLU), which progressively extract and downsample to obtain the hyperprior features $\mathbf{z}$. The hyperprior decoder $h_d$ mirrors the encoder structure, using $5 \times 5$ convolutions and upsampling operations to decode the hyperprior features $\mathbf{z}$ into the distribution parameters (e.g., mean $\mu$ and standard deviation $\sigma$) of the main latent variable, which are then used for subsequent probabilistic modeling and entropy estimation. Each convolutional layer is followed by a ReLU activation, and the number of channels, kernel size, and stride are adapted according to the requirements for downsampling or upsampling.

 \section{Implementation Details of the Variable-length JSCC Encoder and Decoder} \label{appendixF}

This section provides a detailed description of the implementation details of the variable-length JSCC framework.

\begin{figure*}[htbp]
    \centering
    \includegraphics[width=0.95\linewidth]{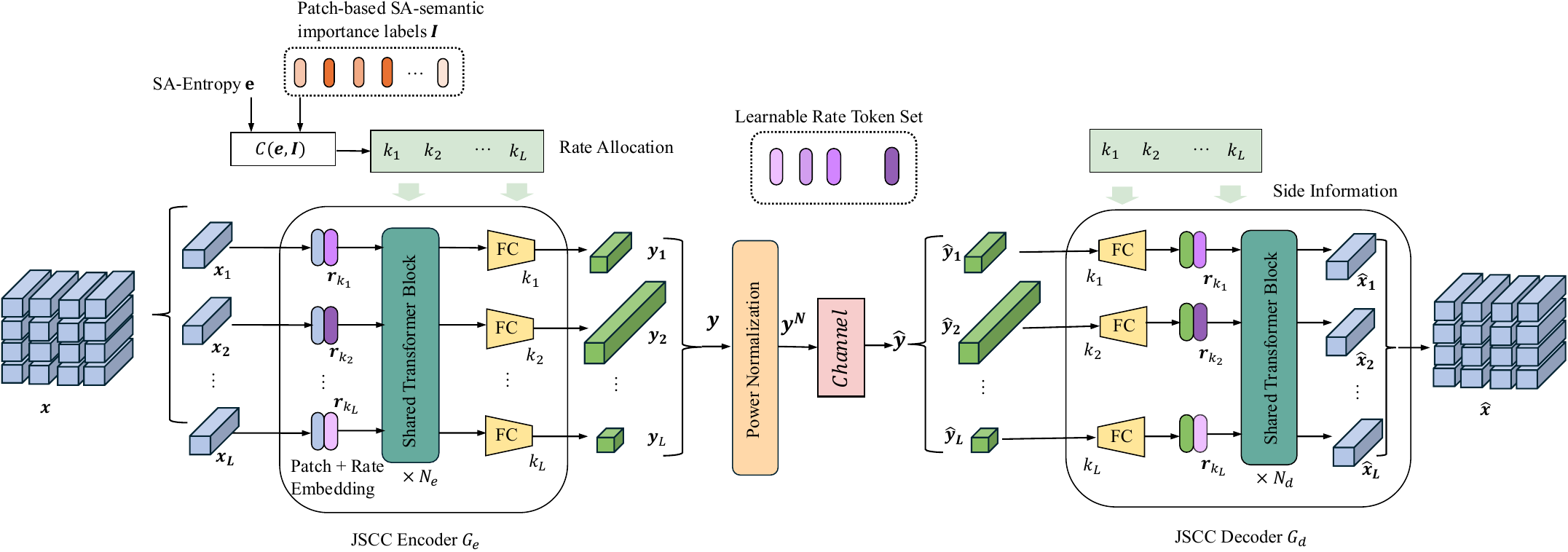}
    \captionsetup{font={small}}
    \caption{The architecture of the variable-length JSCC framework.}
    \label{fig:jscc_arch}
\end{figure*}

As illustrated in Fig. \ref{fig:jscc_arch}, the latent representation sequence $\mathbf{x} = [\mathbf{x}_1, \mathbf{x}_2, \ldots, \mathbf{x}_L]$, produced by the HV vectorization described in the section \ref{sec3}, is fed into the JSCC encoder and mapped into a sequence of variable-length vectors $\mathbf{y} = [\mathbf{y}_1, \mathbf{y}_2, \ldots, \mathbf{y}_L]$. Each $\mathbf{y}_i$ is a $k_i$-dimensional complex-valued signal. As introduced in section \ref{sec3}, to enhance robustness against OOD issues, we formulate $k_i$ according to (\ref{eq:vec_length}). Here, we provide detailed computation of the two terms and the scalar quantization function $Q_0(\cdot)$ in (\ref{eq:vec_length}).

The conventional vector length term $C_1(e_i)$ is given by:
\begin{equation}
C_1(e_i) = \eta \cdot e_i
\label{eq:c1}
\end{equation}
where $\eta$ is a scaling factor. 

The auxiliary vector length term $C_2(I_i)$ is designed to convert the SA semantic importance of each patch labeled from the MLLM into an adaptive adjustment of the output code length. This mechanism aims to mitigate the insufficiency of SA entropy in accurately characterizing the importance of certain OOD objects in real-world applications. $C_2(\cdot)$ is a predefined mapping function, specified as follows:
\begin{equation}
C_2(I_i) = 
\begin{cases}
+\alpha, & I_i = 3 \\
0,       & I_i = 2 \\
-\alpha, & I_i = 1 \\
0,       & I_i = 0 \\
\end{cases}
\label{eq:c2}
\end{equation}
After this mapping, $C_2(I_i)$ represents the additional code length allocated to the $i$-th patch. When the semantic importance $I_i = 3$, an extra $+\alpha$ bandwidth is assigned to the corresponding patch; when $I_i = 2$, no adjustment is made; and when $I_i = 1$, the bandwidth is decreased by $\alpha$. For background patches with $I_i = 0$, the entropy model already allocates very low code length, so no additional adjustment is required.

In Eq.~(1), $Q_0$ denotes a scalar quantizer, which can be expressed as:
\begin{equation}
Q_0 : \mathbb{R} \to V,\quad V = \{ v_1, v_2, \ldots, v_{2^{k_q}} \}
\label{eq:quantizer}
\end{equation}
where $V$ is the quantization set consisting of $2^{k_q}$ integer values ($k_q = 1, 2, \ldots$), and is determined by the scaling factor $\eta$.

Our implementation follows the JSCC framework proposed in \cite{NTSCC}, enabling each $x_i$ to be adaptively mapped to a channel input vector $y_i$ with dimensionality $k_i$. This approach applies the dynamic neural network architecture from \cite{dynamicNN} to the Transformer framework, resulting in a deep JSCC encoder. Such a dynamic network can adaptively adjust its structure and parameters according to different inputs, offering significant advantages in both performance and computational efficiency.

The detailed architecture of the variable-length JSCC encoder is shown in Fig.~\ref{fig:jscc_arch}. The encoder employs $N_e$ shared Transformer modules for feature extraction from input patch embeddings and leverages an adaptive rate allocation mechanism to achieve flexible utilization of channel resources. For each patch embedding $y_i$, it is first concatenated with a learnable rate token $r_{(k_i)}$, forming a joint patch-rate embedding. All learnable rate tokens $r_{(k_i)}$ are drawn from a rate token set $R = \{ r_{(v_1)}, r_{(v_2)}, \ldots, r_{(v_{2^{k_q}})} \}$. Inspired by the positional encoding mechanism in Transformers \cite{transformer}, this design utilizes rate tokens as conditional information, which helps the network to distinguish patch characteristics under different bandwidth allocations.

During feature extraction, each patch embedding is first concatenated with its corresponding rate token and then fed into the shared Transformer blocks. By leveraging self-attention mechanisms, the shared Transformer blocks can infer the required channel bandwidth $k_i$ for each $x_i$. In this way, the Transformer outputs can adaptively reflect the entropy of $y_i$, allowing a subsequent lightweight fully connected layer to efficiently adjust the output dimension to the target $k_i$. This process ultimately generates variable-length channel input vectors $y_i$. This architecture enables the Transformer to dynamically adjust its representations based on the semantic importance and information entropy of each patch, thereby realizing efficient and flexible scenario adaptive JSCC encoding.

At the decoder side, each received $\hat{y}_i$ (which may vary in length) is first projected to a fixed dimension via a fully connected layer and concatenated with the corresponding rate token. The result is then passed through $N_d$ shared Transformer modules to reconstruct patch-level features.

\section{Loss Functions of the Variable-length JSCC Encoder and Decoder}
\label{appendixG}

This section provides a detailed description of the loss function of the variable-length JSCC framework.

The loss function for the variable-length JSCC framework can be interpreted as an optimization of the rate-distortion objective in the transmission scenario, where the key is to balance the overall transmission rate $\|\mathbf{k}\|_1$ and the reconstruction quality of the latent variable $\mathbf{x}$. Specifically, $\|\mathbf{k}\|_1$ denotes the $\ell_1$-norm of $\mathbf{k}$, which is defined as the sum of the absolute values of its elements:
\begin{equation}
\|\mathbf{k}\|_1 = \sum_{i=1}^L |k_i|
\label{eq:l1norm}
\end{equation}

The loss function is defined as follows:
\begin{equation}
\mathcal{L}_{\mathrm{JSCC}} = \lambda \|\mathbf{k}\|_1 + d(\mathbf{x}, \hat{\mathbf{x}})
\label{eq:jscc_loss_basic}
\end{equation}
where $d(\cdot)$ denotes the distortion function, which measures the discrepancy between the latent representation $\mathbf{x}$ and its reconstruction $\hat{\mathbf{x}}$ -- typically using Mean Squared Error (MSE) as the metric. The Lagrange multiplier $\lambda$ balances the trade-off between the total channel bandwidth $\mathbf{k}$ and the end-to-end distortion $d$.

However, given that this task requires different SA semantic importance to be allocated to different patches, we substitute the distortion term with $\mathrm{SAD}(S, \hat{S})$ (see Appendix \ref{appendixB} for more details), which represents the overall image distortion weighted by SA-semantic importance after both the HV networks and the variable-length JSCC codec. Although this approach does not directly minimize the reconstruction distortion of the latent representation $\mathbf{x}$, by directly optimizing the SA semantic importance-weighted distortion between the original and reconstructed images, superior system-level end-to-end reconstruction performance can be achieved. Thus, the final loss function is given by:
\begin{equation}
\mathcal{L}_{\mathrm{JSCC}} = \lambda \|\mathbf{k}\|_1 + \mathrm{SAD}(S, \hat{S})
\label{eq:jscc_loss_sad}
\end{equation}
where $\mathrm{SAD}(S, \hat{S})$ is the SA semantic importance weighted distortion between original image $S$ and its reconstruction $\hat{S}$.

\section{System Training and Overall Loss}
\label{appendixH}

The training process of the entire system is described as follows. It is important to note that, to ensure both faster convergence and training stability, we first pretrain all components involved in the HV network ($f_e$, $f_d$, $h_e$, and $h_d$) under the assumption of a noiseless channel. This pretraining step yields the parameters for the HV network. During this stage, the loss function $\mathcal{L}_{\mathrm{hv}}$ derived in Appendix \ref{appendixD} is used, which effectively optimizes the semantic-aware compression performance of the HV vectorization module.

\textbf{Model Training Details:} The variational encoder $f_e$ and decoder $f_d$ each consist of four stages, containing $N_1 = 1$, $N_2 = 1$, $N_3 = 2$, $N_4 = 6$ Transformer modules, respectively (see Appendix \ref{appendixE} for implementation details). In all experiments, both the JSCC encoder $f_e$ and decoder $f_d$ utilize $N_e = N_d = 4$ Transformer modules, each with 8 multi-head self-attention (MHSA) heads and a channel dimension of 256. The quantization set $V$ contains 16 values uniformly distributed between 16 and 256. During training, the Adam optimizer \cite{adam} is employed with a learning rate set to $10^{-4}$. All experiments are implemented using PyTorch \cite{pytorch}.

Subsequently, during the training of the entire system, in addition to the JSCC loss function \eqref{eq:jscc_loss_sad}, we further include the HV network distortion term $\mathrm{SAD}(S, \hat{S}_{\mathrm{hv}})$ to enhance the stability of convergence and to guarantee the compression performance of the HV network. Therefore, the overall loss function for the system is defined as follows:
\begin{equation}
\mathcal{L}_{\mathrm{overall}} = \lambda \|\mathbf{k}\|_1 + \mathrm{SAD}(S, \hat{S}) + \mathrm{SAD}(S, \hat{S}_{\mathrm{hv}})
\label{eq:overall_loss}
\end{equation}
where $\hat{S}$ denotes the final reconstructed image after both HV and JSCC modules, and $\hat{S}_{\mathrm{hv}}$ denotes the reconstruction using only the HV modules.

\section{Evaluation of the SA-Semantic Importance Identification Module}
\label{sec4sub1}

To assess the consistency between the model’s outputs and human driver decision-making, we introduce human annotation as the evaluation benchmark and compare the importance labels provided by the MLLM with those annotated by human drivers.

Our study uses autonomous driving as the example scenario to measure the MLLM’s semantic understanding under the given task scenario. For the testing dataset, we considered Cityscapes \cite{Cityscapes}, a well-known autonomous driving benchmark that features diverse road environments and a wide range of semantic labels. Given the huge manpower required for manually labeling the importance for objects within the image, it is not possible for tested drivers to annotate the whole Cityscapes dataset. Therefore, we randomly sample 100 images from the Cityscapes dataset to construct the testing dataset. For each image, three experienced drivers are invited to manually annotate the importance of each object in the images, and their consensus serves as the benchmark.

According to the definitions in section \ref{sec2}, importance labels are categorized into four classes: \emph{high}, \emph{medium}, \emph{low}, and \emph{background}. The first three categories correspond to detected objects, while the background category is assigned to undetected regions. To further validate the accuracy of background labeling, special attention is paid during manual annotation to objects that may influence driving decisions but are mistakenly classified as background by the model; such cases are considered as background misclassifications.

The importance labels generated by the MLLM are compared against the human annotations. Experimental results, including the average accuracy for each category and the overall average accuracy, are listed in Table~\ref{tab:acc}.

\begin{table}[htbp]
\captionsetup{font={small}}
\caption{Accuracy of MLLM in SA-semantic importance assignment}
\label{tab:acc}
\begin{center}
\begin{tabular}{|c|c|}
\hline
\textbf{Category} & \textbf{Accuracy (\%)} \\
\hline
High Importance       & 86.10 \\
Medium Importance     & 87.24 \\
Low Importance        & 98.26 \\
Background Importance & 98.00 \\
\hline
\textbf{Overall}      & \textbf{92.57} \\
\hline
\end{tabular}
\end{center}
\end{table}

\begin{figure*}[htbp]
    \centering
    \begin{subfigure}
        \centering
        \includegraphics[width=0.95\linewidth]{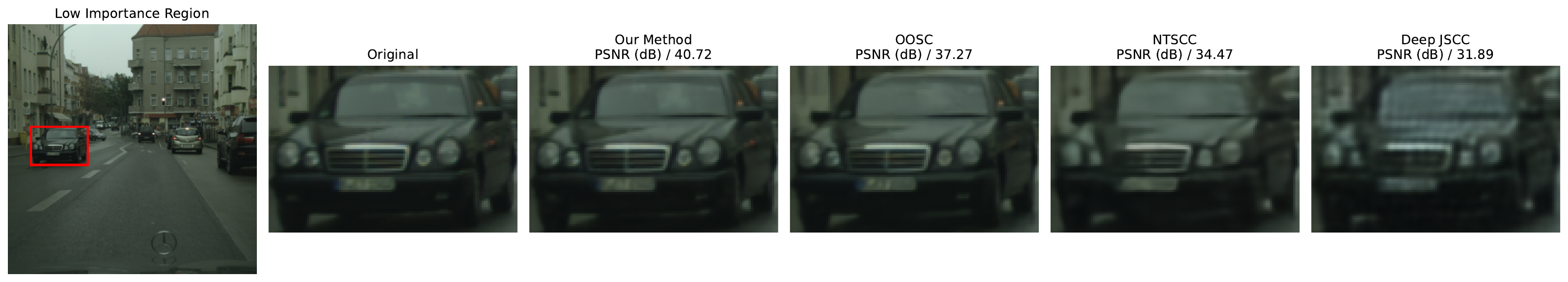}
    \end{subfigure}
    \begin{subfigure}
        \centering
        \includegraphics[width=0.95\linewidth]{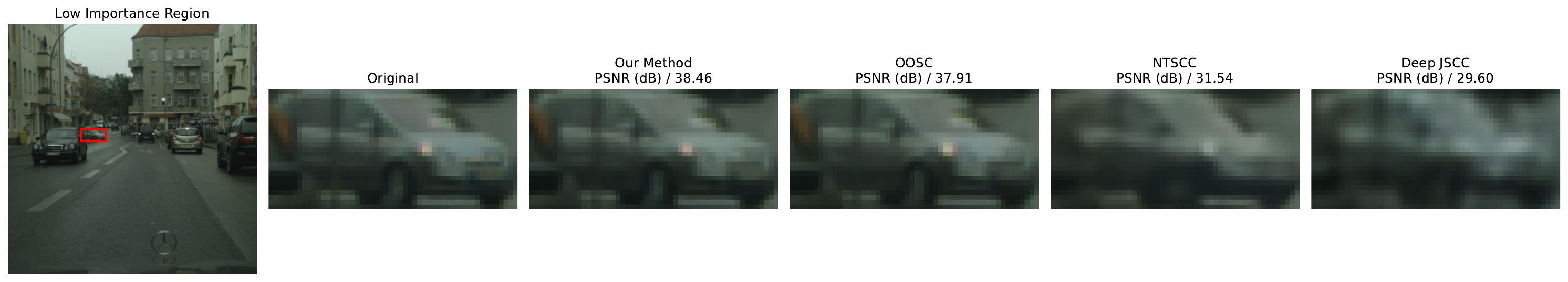}
    \end{subfigure}
    \begin{subfigure}
        \centering
        \includegraphics[width=0.95\linewidth]{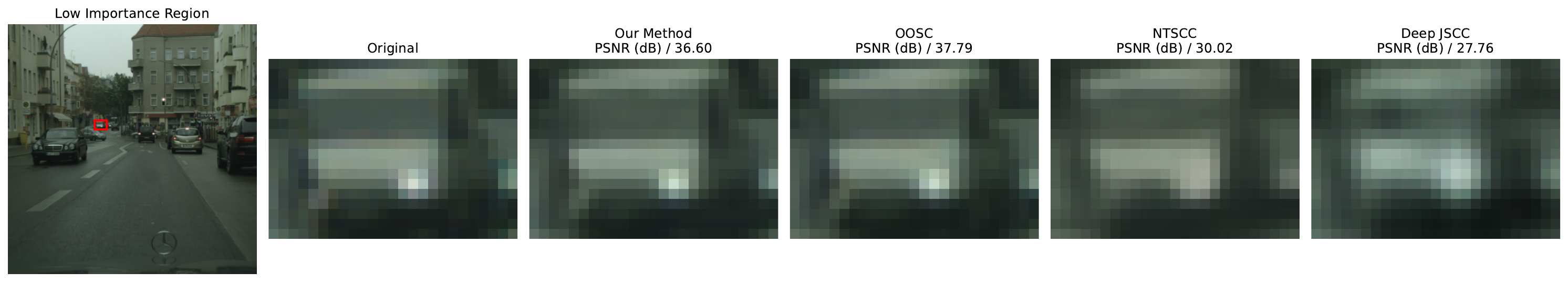}
    \end{subfigure}
    \begin{subfigure}
        \centering
        \includegraphics[width=0.95\linewidth]{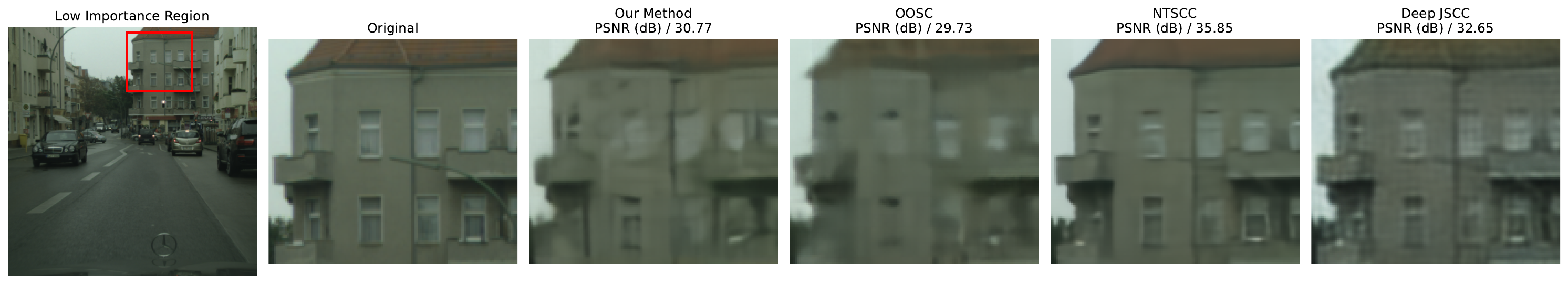}
    \end{subfigure}
    \captionsetup{font={small}}
    \caption{
        Receiver-side reconstruction results and PSNR analysis for different semantic importance regions under various schemes. To ensure fairness in rate comparison, all schemes were configured with nearly identical CBR values for this image, with SA-OOSC achieving the lowest bandwidth utilization (CBR values: 0.0214 for SA-OOSC, 0.0220 for OOSC, 0.0223 for NTSCC, and 0.03 for Deep JSCC). The selected image was encoded by each method at the transmitter, transmitted through an AWGN channel with SNR = 10~dB, and reconstructed at the receiver. The figure presents reconstruction results and corresponding PSNR metrics for four categories of regions: high, medium, low semantic importance, and background. For each category, the left side displays the original image with the region of interest highlighted by a red box, while the right side shows the reconstruction results of SA-OOSC, OOSC, NTSCC, and Deep JSCC in sequence, with the PSNR (dB) for each reconstruction provided below as a quantitative assessment.
    }
    \label{fig:receiver_case_study}
\end{figure*}

In general, our scheme has demonstrated high accuracy in the assignment of SA-semantic importance. The accuracies for the low importance and background categories reach 98.26\% and 98\%, respectively, highlighting the model's strong robustness in background identification and its ability to effectively avoid misclassifying critical objects as background. The accuracies for the high and medium importance categories are slightly lower, primarily due to some high-importance objects being misclassified as medium, and medium-importance objects occasionally being confused with high-importance categories. Additionally, a small number of low-importance objects are not recognized and thus categorized as background, but this has limited impact on the overall performance.

\section{Detailed Case Study on Receiver-Side Reconstruction Performance}
\label{appendixI}

We conducted a detailed case study on the receiver side. The results demonstrate that SA-OOSC significantly outperforms other schemes in reconstructing regions with high semantic importance. For example, in the high-importance region (such as the nearest oncoming vehicle), SA-OOSC achieves a PSNR of 40.72~dB, outperforming OOSC, NTSCC, and Deep JSCC by 3.45~dB, 6.25~dB, and 8.83~dB, respectively, with much clearer detail preservation. In the medium semantic importance region (e.g., a distant vehicle making a turn), SA-OOSC also achieves the highest reconstruction quality (38.46~dB). It provides noticeable advantages in detail preservation, such as faithfully reconstructing the red color of a turning vehicle’s taillight -- a detail missed by OOSC and appearing blurry in both NTSCC and Deep JSCC. In the low semantic importance region (e.g., a more distant parked vehicle), OOSC achieves the best performance (PSNR = 37.79~dB), followed by SA-OOSC (PSNR = 36.60~dB). This is attributed to OOSC allocating uniformly high importance to predefined categories (such as vehicles), without considering the specific position and state of each object within the scenario, resulting in less distinction among the reconstruction performance of regions with different semantic importance. For the background region, both SA-OOSC and OOSC yield low PSNR values, while NTSCC performs best. However, since background regions have negligible impact on downstream decision-making tasks, SA-OOSC deliberately allocates fewer coding resources to these areas to improve overall coding efficiency.

\end{document}